%% file: 0acl_latex.tex
\pdfoutput=1
\documentclass[11pt]{article}
\usepackage{amsmath}
\usepackage{amssymb}
\usepackage{booktabs}
\usepackage{tabularx}
\usepackage{listings}
\usepackage{xcolor}

\usepackage[final]{acl}

\usepackage{times}
\usepackage{latexsym}

\usepackage[T1]{fontenc}
\usepackage[utf8]{inputenc}

\usepackage{microtype}

\usepackage{inconsolata}

\usepackage{graphicx}
\usepackage{caption}

\title{Beyond the Surface: A Solution-Aware Retrieval Model \\for Competition-level Code Generation}

\author{
  \textbf{Shiwen Zhang}\textsuperscript{1,2}, 
  \textbf{Lingxiang Wang}\textsuperscript{1,2}\thanks{\textit{Equally Contribution.}}, 
  \textbf{Hainan Zhang}\textsuperscript{1,2}\thanks{\textit{Corresponding author.}}, \\
  \textbf{Ziwei Wang}\textsuperscript{1,2}, 
  \textbf{Sijia Wen}\textsuperscript{1,2},
  \textbf{Zhiming Zheng}\textsuperscript{1,2} \\
  \textsuperscript{1}Beijing Advanced Innovation Center for Future Blockchain and Privacy Computing \\
  \textsuperscript{2}Institute of Artificial Intelligence, Beihang University, China \\
  \texttt{\{zhangshiwen, wanglingxiang, zhanghainan\}@buaa.edu.cn}
}
\begin{document}
\maketitle
\begin{abstract}
%竞争性编程任务经常利用一段复杂的背景故事进行问题描述，需要深入理解故事背后的解决方案才能通过编程测试，这对大模型无疑是非常有挑战的。如果能够检索到相同解决方案的参考代码，就能有效提高大模型解决竞争性编程问题的能力。不幸的是，当前主流检索模型主要关注文本表面的语义相似性，忽略了文本背后的逻辑相似性，难以直接应用于竞争级代码生成任务。因此，如何构建方案敏感的检索模型召回正确的题目及对应的代码，是一个亟待解决的问题。

%在竞争性编程任务中，问题描述通常嵌入在复杂的叙事背景中，需要对底层解决方案有深入的理解才能成功完成任务。这对当前的代码生成模型提出了重大挑战。检索包含类似解决方案的参考代码可以显著提高此类场景中的模型性能。然而，主流的检索模型主要强调表层语义相似性，忽视了在竞争性代码生成场景中至关重要的深层逻辑和解决方案相似性。因此，设计能够准确识别和检索问题和相应代码的检索模型仍然是竞争性代码生成任务的迫切研究挑战。本文中，我们提出了SolveRank，一种利用合成数据和对比学习为竞争性编程任务训练的方案敏感检索模型。具体地，我们首先利用deepseek-R1模型为每个问题描述生成逻辑相同但描述不同的若干新问题，并使用chatgpt-4o作为判别器筛选解题方案确实相同的新问题。然后，我们利用xxx对比学习方法训练检索模型，以合成数据作为正样本，以BM25检索到的问题和随机采样的问题作为负样本。最后，我们将训练好的检索模型用于测试问题，从训练集中召回相关问题及其参考代码，输入给下游的代码生成模型以完成竞争性编程。在xCodeEval数据集的实验结果显示，SolveRank的检索性能大幅超越传统检索模型（xxx\%以上），编程任务的生成性能（pass@k）超过基线xxx\%，证明了SolveRank确实能够识别到方案相似的编程问题。进一步迁移性实验也证明了SolveRank在其他数据集的有效性。

In competitive programming task, problem statements are often embedded within elaborate narrative backgrounds, requiring deep understanding of the underlying solutions to successfully complete the tasks. Current code generation models primarily focus on token-level semantic modeling, highly susceptible to distractions from irrelevant narrative statements. Inspired by RAG, retrieving reference code with similar solutions may help enhance model performance on difficult problems. However, existing retrieval models also emphasize surface-level semantic similarity, neglecting the deeper solution-level logical similarities that are critical in competitive programming. Therefore, designing ranking models capable of accurately identifying and retrieving problems and corresponding codes remains an urgent research problem in competitive code generation. In this paper, we propose SolveRank, a solution-aware ranking model empowered by synthetic data for competitive programming tasks. Specifically, we leverage the DeepSeek-R1 model to generate logically equivalent but differently phrased new problems, verified by GPT-4o for solution consistency. Then, we train SolveRank with these as positive samples and BM25/random-retrieved problems as negatives. During inference, SolveRank retrieves relevant problems and corresponding code from the corpus to assist a downstream code generator. Experiments on the xCodeEval dataset demonstrate that SolveRank outperforms SOTA ranking methods in precision and recall metrics, and boosts code generation performance for difficult problems.

\end{abstract}

\input{1introduction}
\input{3model}
\input{4experiments}
\input{5conclusion}

\section*{Limitations}
Our experiments are conducted solely on the xCodeEval benchmark, which focuses on competitive programming tasks. The generalizability of our framework to broader code generation domains, such as software engineering tasks or multi-language corpora, remains to be validated in future.

%\section*{Acknowledgments}

\bibliography{custom}

\appendix
\section{Prompt Design}
\label{appendix:prompt}

The following presents the structured prompts used throughout our framework, including (1) generating logically equivalent problems, (2) verifying logical equivalence, and (3) constructing code generation input.

\subsection{Prompt for Logically Equivalent Problem Generation}
\label{appendix:gen_prompt}

%To construct logically equivalent questions for training the retrieval model, we design the following prompt for DeepSeek-R1:caption={Prompt used for generating logically equivalent problems with DeepSeek-R1.},

\begin{lstlisting}[label={lst:prompt_generation}]
You are an algorithm engineer. Given the following problem: {{description}}

Change the background of the question and generate exactly 5 new questions that follow the same logic as the original question, but with different content and background.

Please follow this strict format:
- Output exactly 5 lines.
- Each line must contain exactly one question.
- Do not add any numbering, bullets, or explanations.

To encourage diversity:
- Use a variety of domains or themes such as education, logistics, art, nature, architecture, healthcare, etc.
- Ensure each question uses a different context and vocabulary.

Make sure each generated question is approximately as detailed and long as the original problem. Include any necessary conditions, definitions, and examples if appropriate. Avoid summarizing or oversimplifying the logic.
\end{lstlisting}

\subsection{Prompt for Logical Equivalence Verification}
\label{appendix:verify_prompt}

%To verify whether two questions are logically equivalent, we design the following GPT-4o prompt:caption={Prompt used for logical equivalence verification with GPT-4o.}, 

\begin{lstlisting}[label={lst:prompt_verification}]
Please determine whether the following two questions belong to the same category in terms of modeling logic and algorithmic abstraction. Focus only on their algorithmic modeling structure, core optimization objectives, and typical solution approaches. Ignore the specific real-world background or story.

If the problems are based on the same core abstraction (even with different story settings), answer "Yes". Otherwise, answer "No".

Question A:
{{query}}

Question B:
{{retrieved_question}}

Please answer only "Yes" or "No".
\end{lstlisting}

\subsection{Prompt for Code Generation}
\label{appendix:gen_prompt_code}

%We use the following prompt to guide in-context code generation using the retrieved problem–code pairs:caption={Prompt format used for in-context retrieval-augmented code generation.}, 

\begin{lstlisting}[label={lst:prompt_codegen}]
Write a program in {{lang_cluster}} to solve this programming problem:
Description: {{description}}

{% if retrieved_context %}
Relevant examples (The following examples are selected based on their similarity to the current problem in terms of algorithmic modeling logic and abstraction. They share comparable modeling structures, core optimization objectives, or typical solution strategies. You may ignore the specific application context or surface narrative - focus instead on the underlying algorithmic structure and reasoning process. Use these examples as guidance to help generate code that aligns with the intended problem-solving logic.):
{{retrieved_context}}
{% endif %}

Input Specification: {{input_spec}}
Output Specification: {{output_spec}}

{% for input, output in zip(sample_inputs, sample_outputs) %}
Sample Input:
{{input}}
Sample Output:
{{output}}
{% endfor %}

Notes: {{notes}}

Take input from {{input_from}} and output to {{output_to}}.
Provide the {{lang_cluster}} code without any extra description or tokens. Target code: ||END-of-SRC||
\end{lstlisting}

\section{Case Study}
\subsection{Comparison of Solution-Aware and BM25 Retrieval}
\label{appendix:case study}
\begin{figure*}[t]
    \centering
    \includegraphics[width=0.9\textwidth]{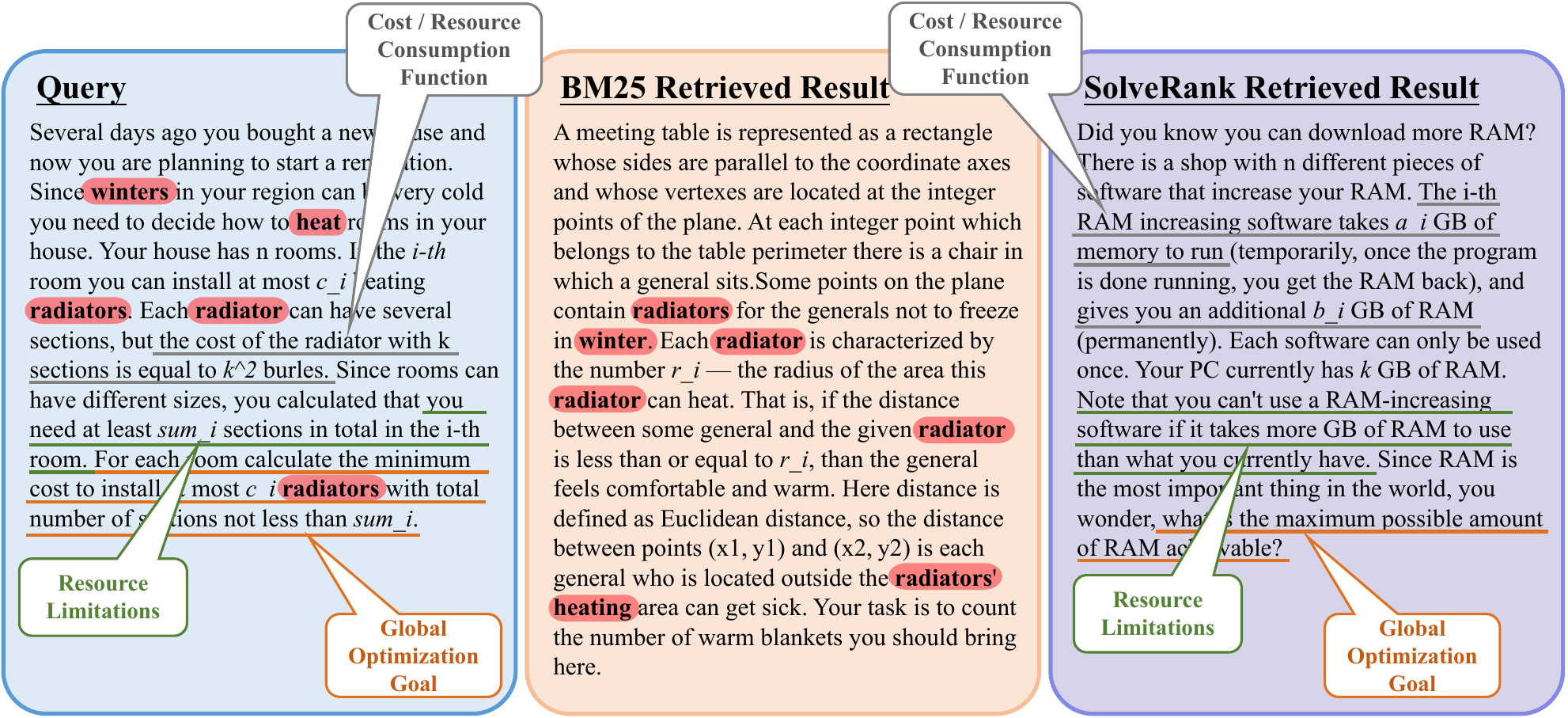}
    \captionsetup{width=0.9\textwidth}
    \caption{An example of retrieval results for a logic-intensive query. BM25 retrieves a problem with similar surface terms such as ``winter'', ``heat'', and ``radiator'', but diverges in algorithmic logic. In contrast, SolveRank retrieves a structurally distinct problem that shares the same underlying optimization goal and reasoning pattern, demonstrating its ability to capture logic-level similarity beyond semantics.}
    \label{fig:retrieval-example}
\end{figure*}

We present a case study in Figure~\ref{fig:retrieval-example} to intuitively illustrate the advantage of \textbf{SolveRank} in retrieving logic-equivalent problems beyond surface semantics, compared to the traditional sparse retriever \textbf{BM25}.

In the given example, the target query describes a resource allocation problem involving heating devices in multiple rooms, where the goal is to install at most \( c_i \) units in each room to achieve a required heat level \( sum_i \), while minimizing a quadratic cost. Although this optimization task is abstract in structure, BM25 is distracted by overlapping terms such as ``radiator'' and ``winter'' in the training set and mistakenly retrieves a problem that focuses on grid coverage via Euclidean distance. The superficial term match misleads the retriever and results in a logically unrelated example that could affect subsequent generation.

In contrast, \textbf{SolveRank} retrieves a problem involving tool selection to increase memory under resource constraints—a task with a completely different surface narrative but an identical optimization structure. Both problems share the same logic: choosing from \( n \) items under a bounded constraint to reach a numeric threshold while minimizing nonlinear cost, which fits a classic dynamic programming pattern. Despite the lack of term overlap, SolveRank successfully captures this deeper alignment.

These results confirm that SolveRank can distinguish structural similarity from surface noise, enabling the retrieval of truly helpful exemplars for code generation, especially when logic alignment is more important than keyword similarity.

\subsection{Medium-difficulty example: tram vs. polar bear sailing}
\label{appendix:example}
We analyze a representative medium-level case. The “tram problem” asks Igor to reach a destination either by walking or by catching a periodically running tram, while the “polar bear sailing problem” requires computing the earliest time to reach a target under wind-driven movement. Since both problems belong to the class of time-constrained shortest reachability, SolveRank considers them logically similar. However, their structures differ: the tram problem is 1D with active boarding choices, whereas the sailing problem is 2D with passive movement:

The tram problem involves a 1D path with periodic opportunities to board.

The sailing problem involves a 2D grid with forced movement according to wind direction.

During code generation, GPT-3.5 misapplied this logic: it failed to model the tram’s periodic arrivals and instead only compared the initial directions. This indicates that for medium-level tasks, directly copying reference examples may mislead weaker models. As model capability improves, however, the retrieved logic can be better adapted to the target problem.

\begin{lstlisting}[label={lst:tram_case},caption={Case study of medium-difficulty task: Tram vs. Polar Bear Sailing.}]
Original Question (Tram):

The tram in Berland goes along a straight line from the point 0 to the point s and back, passing 1 meter per t1 seconds in both directions. It means that the tram is always in the state of uniform rectilinear motion, instantly turning around at points x=0 and x=s.Igor is at the point x1. He should reach the point x2. Igor passes 1 meter per t2 seconds. Your task is to determine the minimum time Igor needs to get from the point x1 to the point x2, if it is known where the tram is and in what direction it goes at the moment Igor comes to the point x1.Igor can enter the tram unlimited number of times at any moment when his and the tram's positions coincide. It is not obligatory that points in which Igor enter and exit the tram are integers. Assume that any boarding and unboarding happens instantly. Igor can move arbitrary along the line (but not faster than 1 meter per t2 seconds). He can also stand at some point for some time.

Reference Question (Polar Bear Sailing):

The polar bears are going fishing. They plan to sail from (sx,sy) to (ex,ey). However, the boat can only sail by wind. At each second, the wind blows in one of these directions: east, south, west or north. Assume the boat is currently at (x,y).  If the wind blows to the east, the boat will move to (x+1,y).  If the wind blows to the south, the boat will move to (x,y-1).  If the wind blows to the west, the boat will move to (x-1,y).  If the wind blows to the north, the boat will move to (x,y+1). Alternatively, they can hold the boat by the anchor. In this case, the boat stays at (x,y). Given the wind direction for t seconds, what is the earliest time they sail to (ex,ey)?
\end{lstlisting}

\input{2relatedwork}

\section{Generalization}
\label{appendix:generalization}
We further evaluate \textsc{SolveRank} on the \textbf{APPS} dataset~\cite{apps}. As shown in Table~\ref{tab:apps-pass1}, \textsc{SolveRank} achieves the highest Pass@1, tied with BM25 . A key difference from xCodeEval (Table~\ref{tab:algo-distribution}) is that \textsc{APPS} problems are generally shorter, with weaker narrative background and a direct focus on input–output specifications. In this setting, semantic retrieval is already sufficient to capture the key information, so the advantage of logic-aware retrieval in filtering narrative noise and aligning solution structures is less apparent.

\begin{table}[t]
\centering
\caption{Results on the \textsc{APPS} dataset (Pass@1).}
\label{tab:apps-pass1}
\begin{tabular}{l c}
\hline
\textbf{Method} & \textbf{Pass@1} \\
\hline
No Retrieval & 24.3\% \\
BM25 & 26.1\% \\
DPR & 24.3\% \\
CodeBERT & 25.2\% \\
SolveRank & 26.1\% \\
\hline
\end{tabular}
\end{table}

\section{Quality of Synthetic Data}

To assess the quality of synthetic positives, we compare them with real problems in terms of prompt length, vocabulary entropy, and average sentence length. As shown in Table~\ref{tab:dist-shift}, all three metrics differ significantly (\(p<0.001\)), confirming that generated problems exhibit diverse phrasing and syntax. Despite this distributional shift, the generated data remain logically consistent and suitable for training solution-level retrieval models.

\begin{table}[!t]
\centering
\caption{Statistical comparison between original and generated problems.}
\label{tab:dist-shift}
\setlength{\tabcolsep}{1pt}  % 可按需要微调列间距
\renewcommand{\arraystretch}{1.05}
\resizebox{\columnwidth}{!}{%
\begin{tabular}{@{}lcccc@{}}
\hline
\textbf{Metric} & \textbf{Original (Avg.)} & \textbf{Generated (Avg.)} & \textbf{\(p\)-value} & \textbf{Significance} \\
\hline
Prompt Length & 275.977 & 190.867 & \(< 0.001\) & Significant \\
Vocabulary Entropy      & 3.863   & 3.635   & \(< 0.001\) & Significant \\
Sentence Length    & 26.331  & 23.667  & \(< 0.001\) & Significant \\
\hline
\end{tabular}%
}
\end{table}

\end{document}

%% file: 1introduction.tex
\section{Introduction}
Recent large language models (LLMs) achieve human-level performance on simple programming tasks~\cite{zheng2023codegeex,wang2025codebc}, but struggle with competitive problems~\cite{li2022competition}. This discrepancy arises because simple programming tasks rely on surface-level instruction following which can be addressed by aligning with human preferences, while competitive problem statements are often embedded within elaborate narrative backgrounds, demanding deeper understanding of the underlying solutions to successfully complete the programming tasks.

\begin{figure}[!t]
    \centering
    \includegraphics[width=0.95\linewidth]{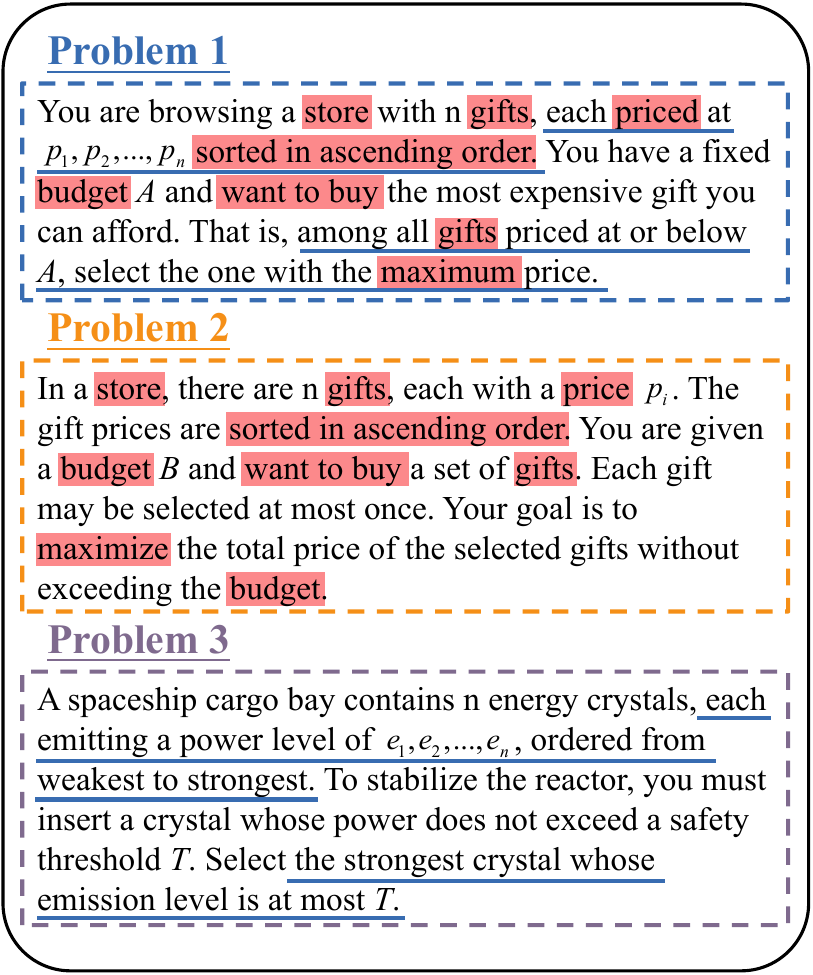}
    \caption{Examples of solution-level logical similar and
semantically similar problems. Problems~1 and Problems~3 differ in surface descriptions but share the same algorithm (binary search). Problem~2 has similar background and vocabulary to Problem~1, yet requires a different solution approach.}
    \label{fig:problem-example}
\end{figure}

Since current code generation models are typically trained using token-level semantic information, they are highly susceptible to being influenced by the narrative problem statements. As illustrated in Figure~\ref{fig:problem-example}, Problems 1 and 2 differ only slightly in wording but require entirely different solutions: Problem 1 is an upper bound search problem in a sorted array, whereas Problem 2 is a 0/1 Knapsack Problem. This semantic similarity can mislead LLMs into generating incorrect solutions by recalling analogous but inappropriate training examples. Therefore, current LLMs still face a significant challenges in solving competitive programming.

\begin{figure}[!t]
    \centering
    \includegraphics[width=1\linewidth]{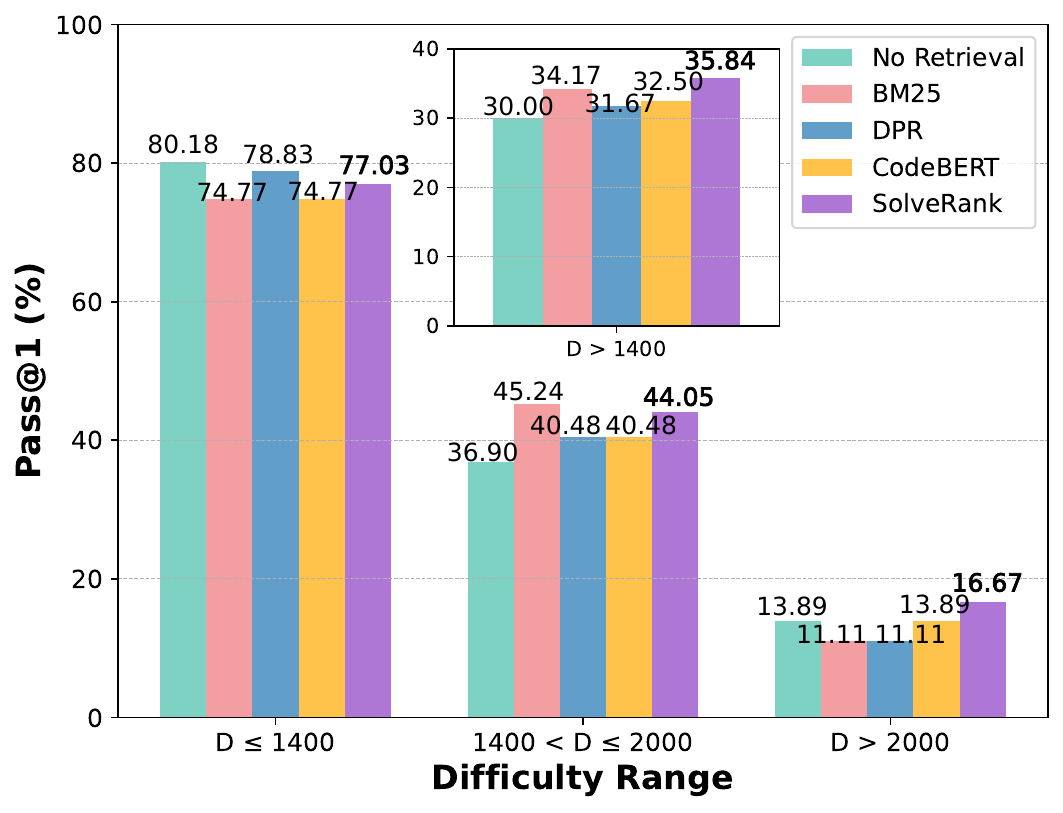}
    \caption{Comparison results of RAG with different ranking methods and without RAG on xCodeEval-python test set. The x-axis is the difficulty score of problems and the y-axis is the pass@1 rate.}
    \label{fig:problem-performance}
\end{figure}

Inspired by Retrieval-Augmented Generation (RAG), reference code that shares the same underlying solution as the current problem may help LLMs generate correct code in competitive programming tasks. As shown in Figure~\ref{fig:problem-performance}, we observe that for simple problems(<=1400), LLMs can directly generate correct code without RAG. But for more challenging problems(>1400), RAG proves to be highly beneficial. Specifically, for difficult problems, retrieving reference code improves the pass@1 rate from 30.00\% to 35.84\%, highlighting the effectiveness of RAG in enhancing LLM performance on competitive programming problems.

However, existing retrieval models primarily focus on surface-level semantic similarity, often retrieving problems with similar wording but lacking deep solution relevance. 
%To assess ranking models for competitive programming, we manually annotated a benchmark of solution-relevant problem pairs (xxx in total) from the xCodeEval-python test set~\footnote{http://xxx}. Results show that state-of-the-art (SOTA) models like BM25 and DPR achieve low Precision@1 scores (13.1\% and 3.86\%, respectively). Furthermore, 
Downstream performance analysis shows that solution-aware retrieval models significantly outperform DPR, achieving a pass@1 of 35.84\% vs. 31.67\% in Figure~\ref{fig:problem-performance}. Notably, inaccurate retrieval in RAG can mislead code generation in competitive programming. Therefore, designing ranking models that can accurately identify and retrieve solution-relevant problems remains an urgent challenge in competitive programming tasks.

In this paper, we propose SolveRank, a solution-aware retrieval model empowered by synthetic data for competitive programming tasks. Specifically, we first leverage the DeepSeek-R1 model to generate multiple synthetic problem statements for current problem that are logically equivalent but differ in surface phrasing. We then use GPT-4o as a discriminator to verify that the generated problems indeed share the same solution with the current problem. Next, we employ contrastive learning to train SolveRank, using the synthetic problems as positive samples, and BM25/random-retrieved real problems as negative. Finally, we apply our SolveRank model to retrieve solution-relevant problems and their reference code from the corpus, which are then provided as context to a downstream code generation model for difficult problems. 

Experiments\footnote{Our code is available at: \href{https://anonymous.4open.science/r/SolveRank-A93B/}{https://anonymous.4open.science/r/SolveRank-A93B/.}} on the xCodeEval~\cite{khan2024xcodeeval} dataset demonstrate that SolveRank significantly outperforms existing retrieval models(about 406\% MRR increment), and improves code generation performance(about 20\% pass@1 increment) for difficult problems, validating the effectiveness of SolveRank on competitive code generation. 

The innovations in this paper are as follows:
\begin{itemize}
    \item We find that RAG is particularly beneficial for solving difficult problems in competitive programming, and solution-aware retrievers outperform semantic-based retrievers. 
    \item We propose a solution-level ranking task focused on assessing whether candidate solutions truly address the query’s intent over surface-level language matching, and release two supporting datasets.
    \item We introduce SolveRank to retrieve relevant solutions to help LLMs generate correct code, significantly improving performance on difficult problems of competitive programming.
\end{itemize}

%最近的大型语言模型（LLM）已经证明了人类在简单编程任务上的表现。然而，它们在复杂的竞争性编程问题上的性能仍然有限。这种差异的产生是因为简单的编程任务通常涉及直接和表面级的指令，使模型只需与人类偏好保持一致即可成功。相比之下，竞争性编程问题通常嵌入在复杂的叙述性语句中，需要对底层解决方案有深入的理解，才能成功实现满足严格时间和空间约束的高效算法。
%由于LLMs的训练过程通常使用的是token级别的语义信息，导致它非常容易受到问题叙述背景信息的影响。如图1所示，问题1的叙述与问题2的叙述看起来非常相似，仅有个别单词不同。但是他们的解决方案却完全不同，问题1是有序集合上的查找问题而问题2是0/1背包问题。因此，如果直接使用LLMs为问题2生成代码，它的生成结果极易受到训练数据中问题1的影响，这显然是错误的。也就是说，当前代码LLMs解决竞争性编程问题仍然面临巨大挑战。
%受到检索增强大模型的启发，如果可以从corpus中检索到与当前问题拥有相同解决方案的参考问题及其对应的参考代码，就有可能帮助LLMs输出竞争性编程问题的正确算法。如图1所示，问题1的解决方案与问题3的解决方案一致，但是他们的问题叙述完全不同，不仔细观察和思考的话，看起来完全不相关。但是将问题3的参考代码与问题1一同给LLMs，它就能够生成通过测试的代码了。此外，我们的实验（see 实验章节）也发现，检索参考代码可以提高竞争性编程的通过率达到xxx%。因此，检索包含相似解决方案的参考代码可以显著提升模型解决竞争性编程的能力。
%不幸的是，当前检索模型也关注的是表面的语义相似性，它是为信息检索系统量身打造的。因此，使用当前排序模型召回的相关问题通常也是看起来相似，忽略了更深层次解决方案级别的逻辑相关。为了衡量检索模型在竞争性编程任务的排序效果，我们针对xCodeEval-python测试集（大约890个问题）上的每个问题，从xCodeEval-python训练集（大约xxx个问题）找到了解决方案相关的真实问题叙述，并进行了人工校验。如表1所示，以bm25、xxx模型为代表的当前检索模型在该标注数据集上的准确率和召回率都不高。值得注意的是，如果检索结果不准确，采用检索增强的方式辅助LLMs进行竞争级编程反而可能误导模型输出错误的代码。因此，设计能够准确识别和检索解决方案相关的问题和代码的排序模型仍然是一个竞争级编程任务亟待解决的问题。

%1、我们提出了解决方案相关性的检索任务。区别于当前的语义排序模型，我们的任务更关注描述背后的解决方案是否相关。2、我们为解决方案相关性任务开源了两个数据集，一个是包含xxx个问题-相关问题对儿的合成数据集，另一个是包含xxx个问题-真实问题-真实代码数据对儿的人工标注的检索性能评价数据集。3、我们提出了SolveRank模型，它通过从corpus中检索解决方案相同的代码，辅助LLMs生成正确的代码，大幅提高了竞争性编程的指标。

%% file: 3model.tex
\section{Method}
\begin{figure*}[!t]
    \centering
    \includegraphics[width=0.9\textwidth,height=0.19\textheight]{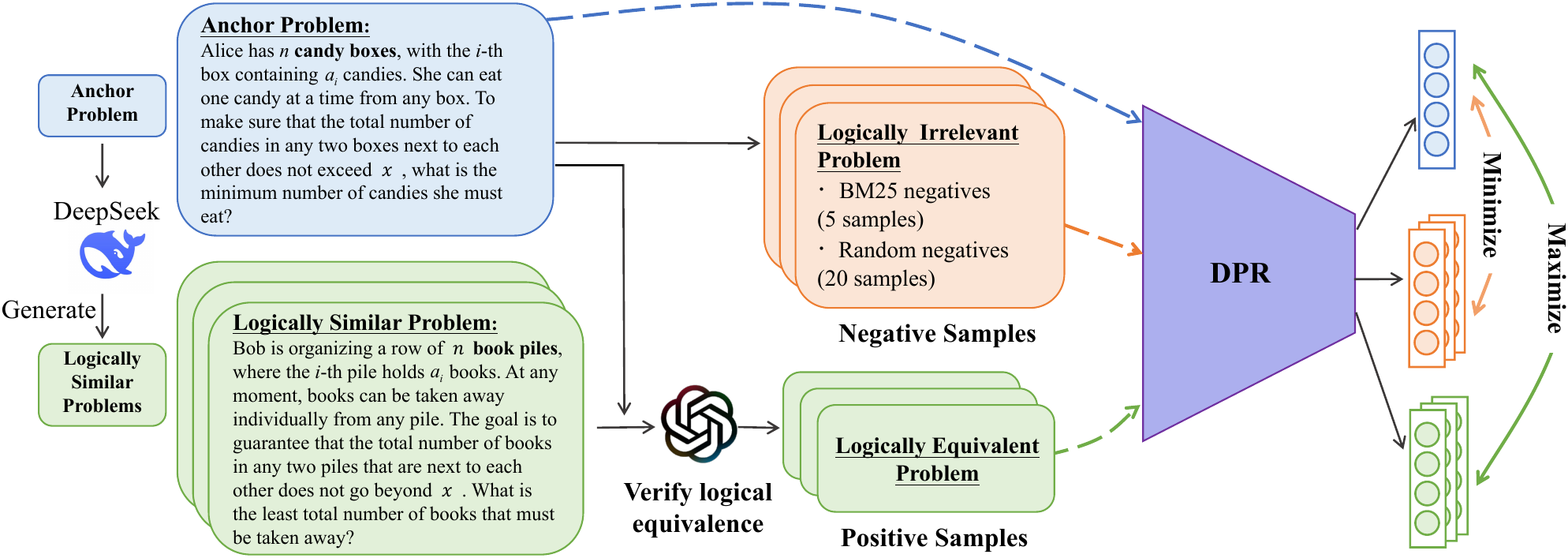}
    \captionsetup{width=0.9\textwidth}
    \caption{Training pipeline for Solution-Aware Retriever \textbf{SolveRank}.}
    \label{fig:retriever-training}
\end{figure*}

The framework of SolveRank consists of three stages, as illustrated in Figure~\ref{fig:retriever-training}. We use the DeepSeek-R1 model to generate logically equivalent but differently phrased problems, which are verified with GPT-4o for consistency. Then, we use these as positive samples to train SolveRank, with BM25/random-retrieved problems as negative ones. During inference, we retrieve relevant problems and code to assist the downstream code generator.

\subsection{Task Definition}
Let \( \mathcal{Q} = \{(q_i, c_i)\}_{i=1}^N \) denote a corpus of natural language programming problems and their corresponding reference codes. Given a target problem \( q^{\prime} \in \mathcal{Q}_{\text{test}}\), the objective of solution-level ranking task is to retrieve a list of problem-code pairs \( \mathcal{R}_q = \{(r_1, c_1), ..., (r_K, c_K)\} \subset \mathcal{Q}_{\text{train}} \) such that each problem \( r_j \) is logically equivalent to \( q^{\prime} \). During code generation, the top-$K$ logically equivalent problem–code pairs $\{(r_j, c_j)\}_{j=1}^K$ are retrieved and concatenated with the target problem $q$ to create an input prompt for code generation models. 

\subsection{Synthetic Data Conduction}
Due to the lack of solution-relevant retrieval data, we construct the training set using synthetic data. Specifically, for each anchor problem \( q \in \mathcal{Q}_{\text{train}} \), we use the DeepSeek-R1 model to generate new problems \( \mathcal{P}_q = \{q_i^+\}_{i=1}^{5} \). The prompt (see Appendix~\ref{appendix:gen_prompt}) is crafted to preserve the original solution logic while encouraging diversity in the problem background.

To ensure true logical equivalence, we apply GPT-4o as an automatic verifier. For each generated variant \( q_i^+ \in \mathcal{P}_q \), GPT-4o is prompted to assess whether \( \text{Logic}(q^+) \equiv \text{Logic}(q) \), focusing strictly on algorithm class and solution decomposition while ignoring superficial narrative or vocabulary differences(see in Appendix~\ref{appendix:verify_prompt}).A statistical comparison of synthetic and original problems is provided in Appendix~\ref{tab:dist-shift}.

\subsection{Solution-Aware Retriever}
\label{subsec:retriever}
%We adopt a dense retrieval model DPR~\cite{karpukhin2020dense} as the backbone, consisting of a query encoder \( E_Q(\cdot) \) and a passage encoder \( E_P(\cdot) \). The similarity between the query and candidate problem is:
%\begin{equation}
%\text{sim}(q, r) = E_Q(q)^\top E_P(r).
%\label{eq:similarity}
%\end{equation}

We adopt a contrastive learning approach to train a DPR model~\cite{karpukhin2020dense}. The \textbf{positive samples} are drawn from the synthetic dataset \( \mathcal{P}_q \), while the \textbf{negative samples} \( \mathcal{N}_q = \{q_j^-\}_{j=1}^{25} \) consist of the top-5 retrieved by BM25 and 20 randomly sampled problems from the training corpus.

We use the \textbf{InfoNCE} loss to encourage the encoder to bring logic-equivalent problems closer in the embedding space while pushing apart logical distractors. The loss is defined as:
% \begin{align}
%     &\mathcal{L} = -\log \frac{\exp\left( \frac{\text{sim}(q, q^+)}{\tau} \right)}{\exp\left( \frac{\text{sim}(q, q^+)}{\tau} \right) + \sum\limits_{q^- \in \mathcal{N}_q} \exp\left( \frac{\text{sim}(q, q^-)}{\tau} \right)}, \\
%     &\text{sim}(q, r) = E_Q(q)^\top E_P(r),
% \label{eq:loss}
% \end{align}

\begin{align}
    &\mathcal{L} = -\log \frac{\exp\left( \frac{\text{sim}(q, q^+)}{\tau} \right)}{\exp\left( \frac{\text{sim}(q, q^+)}{\tau} \right) + \sum\limits_{q^- \in \mathcal{N}_q} \exp\left( \frac{\text{sim}(q, q^-)}{\tau} \right)}, \\
    &\hspace*{4.5em} \text{sim}(q, r) = E_Q(q)^\top E_P(r),
\label{eq:loss}
\end{align}
where \( \tau \) is a hyperparameter,  \( E_Q(\cdot) \) and \( E_P(\cdot) \) are query and passage encoders from DPR model.

After that, given a target problem \( q^{\prime} \) and its top-\( K \) retrieved candidates \( \{r_1, r_2, ..., r_K\} \), we verify whether each \( r_j \) satisfies logical equivalence with \( q \) with GPT-4o judgment(see in Appendix~\ref{appendix:verify_prompt}).

\subsection{Retrieval-Augmented Code Generation}
Given the top-\( K \) problem–code pairs \( \{(r_j, c_j)\}_{j=1}^K \) retained for the target problem \( q^{\prime} \) , we concatenate the verified examples and the target problem into a single input, formatted as:
\[
\text{Prompt}(q) = \text{Concat} \left( \{(r_1, c_1), \ldots, (r_K, c_K)\}, \; q \right).
\]
We use large language models to generate Python code in an autoregressive manner. The model operates in a zero-shot setting, generating the code until an end-of-function token is produced or a maximum length is reached.

%% file: 4experiments.tex
\section{Experiments}
\subsection{Experimental Setups}

\paragraph{Dataset}
We use the \textbf{xCodeEval} benchmark for experiments, which includes competitive programming problems in areas like dynamic programming, graph traversal, greedy algorithms, and simulation. The official \textbf{test set} of the NL-Code Retrieval task is used for evaluation, while the training set serves as the retrieval corpus. Since the \texttt{ExecEval} platform of xCodeEval only supports problems from the \texttt{program\_synthesis} subset for functional evaluation, we filter the NL-Code Retrieval test set to keep 342 suitable problems.

\paragraph{Baselines}
We compare SolveRank with three SOTA ranking methods: BM25~\cite{robertson2009probabilistic}, CodeBERT~\cite{feng2020codebert} , DPR~\cite{karpukhin2020dense} and ReACC~\cite{wan2022reacc}.
\paragraph{Evaluation Metrics}
We use Pass@1 as the primary metric for competition-level code generation, measuring the proportion of problems where the top-1 generated code passes all test cases via ExecEval platform. For the solution-level ranking task, we use Precision(P@K), Recall(R@K) and MRR to evaluate the model performance.

\paragraph{Implementation Details}
We use GPT-4o and GPT-3.5 as the code generation models in zero-shot inference mode. SolveRank is trained on a dual-GPU server (NVIDIA RTX A6000) for 10 epochs using a batch size of 4 and a learning rate of $3 \times 10^{-5}$, under CUDA 12.5.

% \begin{table}[!t]
% \centering
% \scriptsize
% \setlength{\tabcolsep}{4.5pt}

% \begin{minipage}{\linewidth}
% \centering
% \begin{tabular}{lccc}
% \toprule
% \textbf{Method} & \textbf{$D \leq 1400$} & \textbf{$1400 < D \leq 2000$} & \textbf{$D > 2000$} \\
% \midrule
% No Retrieval   & \textbf{49.77} & 10.71 & 5.56 \\
% BM25           & 38.74 & 14.29 & 8.33 \\
% DPR            & 45.50 & 11.90 & 5.56 \\
% CodeBERT       & 41.18 & \textbf{17.86} & \textbf{11.11} \\ \hline
% \textbf{SolveRank} & 40.54 & 13.10 & \textbf{11.11} \\
% \bottomrule
% \end{tabular}
% \caption*{\textbf{(a)}}
% \end{minipage}

% \vspace{0.5em}

% \begin{minipage}{\linewidth}
% \centering
% \begin{tabular}{lccc}
% \toprule
% \textbf{Method} & \textbf{$D \leq 1400$} & \textbf{$1400 < D \leq 2000$} & \textbf{$D > 2000$} \\
% \midrule
% No Retrieval   & \textbf{80.18} & 36.90 & 13.89 \\
% BM25           & 74.77 & \textbf{45.24} & 11.11 \\
% DPR            & 78.83 & 40.48 & 11.11 \\
% CodeBERT       & 74.77 & 40.48 & 13.89 \\ \hline
% \textbf{SolveRank} & 77.03 & \textbf{44.05} & \textbf{16.67} \\
% \bottomrule
% \end{tabular}
% \caption*{\textbf{(b)}.}
% \end{minipage}

% \caption{Pass@1 (\%) for different difficulty levels on xCodeEval using GPT-3.5(a) and GPT-4o(b). $D$ denotes the official difficulty score.}
% \label{tab:pass1-both}
% \end{table}

\begin{table}[!t]
\centering
\scriptsize
\setlength{\tabcolsep}{4.5pt}
\caption{Pass@1 (\%) for different difficulty levels on xCodeEval using GPT-3.5(a) and GPT-4o(b). $D$ denotes the official difficulty score.}
\label{tab:pass1-both}

% ----------- (a) -----------
\begin{minipage}{\linewidth}
\centering
\begin{tabular}{lccc}
\toprule
\textbf{Method} & \textbf{$D \leq 1400$} & \textbf{$1400 < D \leq 2000$} & \textbf{$D > 2000$} \\
\midrule
No Retrieval   & \textbf{49.77} & 10.71 & 5.56 \\
Random           & 38.29 & 10.71 & 5.56 \\
BM25           & 38.74 & 14.29 & 8.33 \\
DPR            & 45.50 & 11.90 & 5.56 \\
ReACC      & 43.69 & 13.10 & 5.56 \\
CodeBERT       & 41.18 & \textbf{17.86} & \textbf{11.11} \\
\textbf{SolveRank} & 40.54 & 13.10 & \textbf{11.11} \\
\bottomrule
\end{tabular}
\vspace{0.3em} % 更紧凑
\par{\textbf{(a)}}
\end{minipage}

\vspace{0.3em} % 两张表之间紧凑

% ----------- (b) -----------
\begin{minipage}{\linewidth}
\centering
\begin{tabular}{lccc}
\toprule
\textbf{Method} & \textbf{$D \leq 1400$} & \textbf{$1400 < D \leq 2000$} & \textbf{$D > 2000$} \\
\midrule
No Retrieval   & \textbf{80.18} & 36.90 & 13.89 \\
Random           & 72.52 & 42.86 & 11.11 \\
BM25           & 74.77 & \textbf{45.24} & 11.11 \\
DPR            & 78.83 & 40.48 & 11.11 \\
ReACC           & 72.52 & 38.10 & 13.04 \\
CodeBERT       & 74.77 & 40.48 & 13.89 \\
\textbf{SolveRank} & 77.03 & \textbf{44.05} & \textbf{16.67} \\
\bottomrule
\end{tabular}
\vspace{0.3em}
\par{\textbf{(b)}}
\end{minipage}

\end{table}

\begin{table}[t]
\centering
\scriptsize
\caption{Ranking performance by retrieving synthetic solution-relevant problems from xCodeEval-python.}
\label{tab:retrieval-discussion}
\setlength{\tabcolsep}{4pt}
\begin{tabular}{lcccccccccc}
\toprule
\textbf{Method} & \textbf{P@1} & \textbf{R@1} & \textbf{P@3} & \textbf{R@3} & \textbf{P@5} & \textbf{R@5} & \textbf{MRR} \\
\midrule
BM25            & 0.131 & 0.026 & 0.198 & 0.068 & 0.240 & 0.103 & 0.186 \\
DPR & 0.039 & 0.008 & 0.059 & 0.020 & 0.069 & 0.030 & 0.057 \\
ReACC & 0.027 & 0.005 & 0.064 & 0.016 & 0.083 & 0.024 & 0.057 \\
CodeBERT        & 0.096 & 0.019 & 0.167 & 0.048 & 0.193 & 0.066  & 0.147 \\ \hline
SolveRank       & \textbf{0.682} & \textbf{0.136} & \textbf{0.808} & \textbf{0.385} & \textbf{0.842} & \textbf{0.593} & \textbf{0.755} \\
\bottomrule
\end{tabular}

\end{table}

\subsection{Main Results}

The study evaluates Pass@1 performance across three difficulty levels (Easy<=1400, 1400 < Medium <=2000, and Hard>2000) on the xCodeEval dataset using GPT-3.5-turbo and GPT-4o, as shown in Table~\ref{tab:pass1-both}. The results show that for easy problems, retrieval offers no significant improvement, sometimes even decreasing performance. This suggests that LLMs can already solve simple problems effectively without additional guidance. In contrast, for medium and hard tasks, all the retrieval methods enhance performance, indicating that RAG is helpful for more complex problems. 

For easy problems, all retrieval models surpass the Random baseline, showing they can capture some relevance. However, compared to the No Retrieval setting, all methods perform worse, regardless of whether the retrieved examples are semantically similar, logically aligned, or randomly sampled. This suggests that the base model is already sufficient to solve simple tasks independently, and adding reference may introduce distractions and degrade performance.

For medium problems, SolveRank may perform worse as its logic-aware retrieval focuses on complex, deep examples, which can introduce unnecessary abstraction and cognitive load for simpler tasks. A more detailed analysis of this phenomenon will be presented in Section~\ref{further analysis}.But for hard problems, SolveRank outperforms all other baselines, especially with GPT-4o. SolveRank yields a Pass@1 of 16.67\% with GPT-4o and 11.11\% with GPT-3.5, while other methods (e.g., DPR, BM25, and ReACC) offer little to no improvement compared to no retrieval. These results show that logic-equivalent examples retrieved by SolveRank aid in solving complex problems, highlighting the importance of structural alignment and deep solution in retrieval-augmented code generation.Further evaluation on the APPS dataset is provided in Appendix~\ref{appendix:generalization}.

We evaluate the ranking performance by comparing SolveRank with SOTA ranking baselines, using the retrievability of synthetic solution-relevant problems from the xCodeEval-Python training dataset as the evaluation criterion. From Table~\ref{tab:retrieval-discussion}, we can see that SolveRank outperforms all baselines, achieving a P@1 of 0.682 and an MRR of 0.755, while BM25, ReACC, and CodeBERT have much lower MRRs (0.186, 0.057 and 0.147, respectively). BM25 focuses on lexical overlap, leading to irrelevant results, while CodeBERT struggles with solution-level structure. ReACC focuses on surface semantic similarity through code transformations and API usage, which is less effective for competitive programming tasks.SolveRank, through contrastive learning, captures algorithmic alignment and reasoning logic, making it more effective for competitive programming.To further illustrate this distinction, a case study is provided in Appendix~\ref{appendix:case study}.

\subsection{Further Analysis}
\label{further analysis}
\begin{table}[t]
\centering
\footnotesize
\setlength{\tabcolsep}{6pt} % 需要更紧凑可再调小
\caption{Algorithm distribution across difficulty levels in xCodeEval.}
\label{tab:algo-distribution}
\resizebox{\linewidth}{!}{%
\begin{tabular}{lccc}
\toprule
\textbf{Algorithm Tag} & $\leq 1400$ & $1400$--$2000$ & $\geq 2000$ \\
\midrule
implementation         & 146 & 26 & 18 \\
math                   & 91  & 50 & 18 \\
brute force            & 62  & 25 & 2  \\
greedy                 & 50  & 8  & 4  \\
dp                     & 17  & 19 & 18 \\
constructive algorithms& 15  & 29 & 1  \\
number theory          & 14  & 11 & 4  \\
strings                & 14  & 1  & 0  \\
sortings               & 11  & 1  & 0  \\
binary search          & 8   & 5  & 5  \\
bitmasks               & 5   & 6  & 2  \\
combinatorics          & 4   & 12 & 2  \\
probabilities          & 4   & 6  & 0  \\
shortest paths         & 2   & 0  & 3  \\
graphs                 & 2   & 4  & 4  \\
dfs and similar        & 2   & 8  & 3  \\
geometry               & 1   & 6  & 1  \\
games                  & 1   & 6  & 6  \\
matrices               & 1   & 2  & 3  \\
two pointers           & 1   & 0  & 0  \\
expression parsing     & 1   & 0  & 0  \\
data structures        & 0   & 12 & 1  \\
divide and conquer     & 0   & 2  & 3  \\
flows                  & 0   & 0  & 13 \\
fft                    & 0   & 0  & 1  \\
trees                  & 0   & 0  & 3  \\
graph matchings        & 0   & 1  & 2  \\
meet-in-the-middle     & 0   & 1  & 0  \\
string suffix structures & 0 & 0  & 1  \\
\midrule
\textbf{total}         & 452 & 241 & 118 \\
\bottomrule
\end{tabular}%
}
\end{table}

\begin{table}[t]
\centering
\footnotesize 
\caption{SolveRank error distribution by problem type.}
\label{tab:error-distribution}
\setlength{\tabcolsep}{6pt}
\begin{tabular}{lcc}
\toprule
\textbf{Algorithm Tag} & \textbf{GPT-3.5} & \textbf{GPT-4o} \\
\midrule
math                    & 9 & 7 \\
implementation          & 5 & 6 \\
greedy                  & 5 & 4 \\
brute force             & 4 & 3 \\
dfs and similar         & 4 & 2 \\
combinatorics           & 3 & 0 \\
dp                      & 3 & 0 \\
graphs                  & 3 & 0 \\
number theory           & 2 & 2 \\
dsu                     & 2 & 0 \\
constructive algorithms & 1 & 0 \\
binary search           & 1 & 1 \\
geometry                & 1 & 2 \\
bitmasks                & 1 & 0 \\
data structures         & 0 & 1 \\
\bottomrule
\end{tabular}
\end{table}

While SolveRank demonstrates strong performance on difficult problems, we observe that its advantage is less prominent on medium-difficulty tasks. To better understand this phenomenon, we conduct a detailed analysis of algorithm distribution, error cases, and model behaviors.

As shown in Table~\ref{tab:algo-distribution}, easy problems ($D \leq 1400$) are dominated by implementation, math, brute force, greedy.Medium problems ($1400 < D \leq 2000$) are still concentrated on math, implementation, and brute force.Hard problems ($D > 2000$) exhibit a more balanced distribution and contain many algorithm classes that are rarely present in lower levels, such as flows, FFT, trees, graph matchings, string suffix structures.

We further examine the cases where SolveRank fails but at least one semantic retriever (BM25, DPR, ReACC, or CodeBERT) succeeds. Table~\ref{tab:error-distribution} shows that most errors occur in math, implementation, greedy, and brute force categories. This highlights a limitation of solution-aware retrieval:

For math problems, which are already highly abstract, additional reference problems provide limited benefit; success depends more on the code generator’s inherent mathematical reasoning ability.For implementation and greedy problems, semantic retrievers often identify problems with similar contexts or scenarios, which can better guide the generator to simulate processes correctly.
In contrast, SolveRank excels at identifying deep structural and algorithmic similarities, which are more critical in complex, high-difficulty tasks.To illustrate this, we analyze a representative medium-level case in Appendix~\ref{appendix:example}.

%% file: 5conclusion.tex
\section{Conclusion}
In competitive programming, understanding problem-solving logic is crucial. Current code generation models focus on surface-level semantics, which often fail on complex problems. This paper introduces SolveRank, a solution-aware ranking model that uses synthetic data to improve code generation performance. The model outperforms semantic-based retrievers and introduces a solution-level ranking task. Future work will explore the reinforcement learning for ranking improvement.

%% file: 2relatedwork.tex
\section{Related Work}

%\subsection{Retrieval-Augmented Generation (RAG)}

%Retrieval-Augmented Generation (RAG)~\cite{siriwardhana2022improving} is a widely adopted framework that enhances language models by incorporating retrieved context from an external corpus. A typical RAG pipeline consists of two components: a \textbf{retriever}, which selects the top-$k$ relevant passages (e.g., documents, examples, code snippets) given a query; and a \textbf{generator}, which conditions on both the original input and the retrieved content to produce output. RAG has shown substantial benefits in open-domain question answering, fact verification, and long-context generation tasks, improving factuality, interpretability, and generalization.

%Recent RAG extensions also focus on in-context learning and retrieval-augmented pretraining, showing that properly selected retrieval inputs can significantly reduce hallucination and improve sample efficiency. However, the success of RAG heavily depends on the quality and relevance of the retrieved examples. While semantic similarity is often sufficient in natural language tasks, it may fall short in domains such as programming, where deeper structural alignment is required.

\subsection{Retrieval-Augmented Code Generation}

The RAG paradigm has been increasingly adopted in natural language to code (NL2Code) generation. In this setting, the model retrieves code-related knowledge (e.g., semantically similar problems or code snippets) and uses it as context to generate functionally correct programs. \textbf{BM25}~\cite{rosa2021bm25} is a classical sparse retrieval method based on term frequency and inverse document frequency (TF-IDF). Despite its simplicity, BM25 is commonly used as a baseline due to its high precision for short queries. \textbf{CodeBERT-Retrieval}~\cite{zhang2020codebert} leverages the CodeBERT encoder to encode NL-code pairs, building a bi-encoder retriever to retrieve semantically similar problems based on cosine similarity of embeddings. \textbf{UniXcoder-Retrieval}~\cite{yin2021unixcoder} extends CodeBERT with unified cross-modal representations, integrating NL, AST, and code tokens for richer retrieval. It supports both encoder-only and encoder-decoder settings and has shown better performance in code-related retrieval tasks. \textbf{ReACC}~\cite{wan2022reacc} proposes retrieval-augmented contrastive training. It retrieves code snippets as positive contexts during training, thereby improving generalization on unseen NL2Code samples.

These methods have demonstrated improvements on CodeSearchNet~\cite{husain2019codesearchnet}, CoNaLa~\cite{liu2019conala}, and HumanEval~\cite{chen2021humaneval}. However, their retrieval strategies are predominantly based on surface-level similarity, which overlooks deeper logic algorithms.

\subsection{Logical Reasoning in Programming}
Logical reasoning is a critical capability for solving structured programming problems that require deeper understanding beyond surface-level semantics. 
Recent works have proposed combining large language models (LLMs) with symbolic reasoning systems to address this limitation. Logic-LM~\cite{pan2023logiclm} enhances the faithfulness of reasoning by translating natural language queries into formal logic representations and solving them using symbolic solvers. Similarly, DSR-LM~\cite{zhang2023dsrlm} introduces differentiable symbolic reasoning modules into LLMs, enabling fine-grained rule induction and significantly improving performance on logic-intensive tasks. In the context of knowledge-grounded reasoning, LACT~\cite{xia2024lact} applies a logic-aware curriculum tuning strategy to improve the model's ability to perform multi-hop and inductive reasoning over knowledge graphs. This approach highlights the importance of reasoning difficulty control and progressive learning in complex code understanding scenarios. 
Moreover, coupling LLMs with logic programming frameworks such as answer set programming has shown promising generalization capabilities~\cite{yang2023logicprog}. This hybrid paradigm allows models to abstract structural logic patterns from textual descriptions and execute them robustly, even in previously unseen settings.

Existing methods often rely on surface-level similarity when retrieving examples for code generation. In contrast, we propose \textbf{SolveRank} to capture deeper solution-level similarities, thus offering more relevant and generalizable retrievals for downstream code generation tasks.